\documentclass[12pt,onecolumn]{revtex4}
\usepackage{amssymb,epsf}
\begin{document}

\title{Abelian Higgs Hair for a Static Charged Black String}
\author{M. H. Dehghani}\email{dehghani@physics.susc.ac.ir}
\address{Department of Physics and
Biruni  Observatory, College of Sciences, Shiraz
University, Shiraz 71454, Iran\\and\\
Institute for Studies in Theoretical Physics and Mathematics
(IPM), P.O. Box 19395-5531, Tehran, Iran}
\author{T. Jalali}\email{jalali@persiangulfu.ac.ir}
\address{College of Sciences, Persian Gulf University,
Bushehr, Iran}
\begin{abstract}
We study the problem of vortex solutions in the background of an
electrically charged black string. We show numerically that the
Abelian Higgs field equations in the background of a
four-dimensional black string have vortex solutions. These
solutions which have axial symmetry, show that the black string
can support the Abelian Higgs field as hair. This situation holds
also in the case of the extremal black string. We also consider
the self-gravity of the Abelian Higgs field and show that the
effect of the vortex is to induce a deficit angle in the metric
under consideration.
\end{abstract}

\maketitle


\section{Introduction}

The conjecture that after the gravitational collapse of the matter
field the resultant black hole is characterized by at most its
electromagnetic charge, mass, and angular momentum is known as the
classical no-hair conjecture and was first proposed by Ruffini and
Wheeler \cite{Ruf}. Nowadays we are faced with the discovery of
black hole solutions in many theories in which Einstein's equation
is coupled with some self interacting matter fields, and therefore
this conjecture needs more investigation. In certain special cases
the conjecture has been verified. For example, a scalar field
minimally coupled to gravity in asymptotically flat spacetimes
cannot provide hair for the black hole \cite{Sud}. But this
conjecture cannot extend to all forms of matter fields. It is
known that some long range Yang-Mills quantum hair could be
painted on the black holes \cite{Eli}. Explicit calculations have
also been carried out which verify the existence of a long range
Nielsen-Olesen vortex solution as stable hair for a Schwarzschild
black hole in four dimensions \cite{Achu}. Of course one may note
that this situation falls outside the scope of the classical
no-hair theorem due to the nontrivial topology of the string
configuration.

Recently some effort has been made to extend these ideas to the
case of (anti-)de Sitter spacetimes. While a scalar field
minimally coupled to gravity in asymptotically de Sitter
spacetimes cannot provide hair for the black hole \cite{Torii1},
it has been shown that in asymptotically AdS spacetime there is a
hairy black hole solution \cite{Torii2}. Also, in Ref. \cite{Eli},
it was shown that there exists a solution to the $SU(2)$
Einstein-Yang-Mills equations which describes a stable Yang-Mills
hairy black hole that is asymptotically AdS. In addition the idea
of Nielsen-Olesen vortices has been extended to the case of
asymptotically (anti)-de Sitter spacetimes \cite{Deh1,Ghez1}. The
investigation of Nielsen-Olesen vortices in the background of
charged black holes was done in \cite{Cha1}-\cite{Bon2}. More
recently the stability of the Abelian Higgs field in
AdS-Schwarzschild and Kerr-AdS backgrounds has been investigated
and it has been shown that these asymptotically AdS black holes
can support an Abelian Higgs field as hair \cite{Deh2,Ghez2}.

Motivated by these subjects, in this paper we investigate possible
solutions of the Abelian-Higgs field equations in a
four-dimensional charged black string background\cite{Lem,Deh3}.
While an analytical solution to these equations appears to be
intractable, we confirm by numerical calculation that charged
black string could support an Abelian Higgs field as its hair.

In Sec. \ref{Vor}, we consider the Abelian Higgs field equations
in the background of a charged black string. Section \ref{Num} is
devoted to the numerical solutions of the field equations for
different values of charge and mass parameters and winding number.
In Sec. \ref{Self}, by studying the behavior of the Abelian Higgs
field energy-momentum tensor, we find the effect of the vortex
self-gravity on the charged black string background. We give some
closing remarks in the final section.

\section{Abelian Higgs Vortex in the Presence of a Charged Black
String}\label{Vor}

In this section we study the Nielsen-Olesen equations for an
Abelian Higgs vortex in the presence of an electrically charged
black string background. The system may be described by the action
\begin{equation}
I=I_G+I_H.  \label{totact}
\end{equation}
The first term of Eq. (\ref{totact}) is the gravitational action
of four-dimensional asymptotically anti-de Sitter spacetimes in
the presence of an electromagnetic field given by
\begin{equation}
I_G=-\frac 1{16\pi }\int_{\mathcal{M}}d^4x\sqrt{-g}\left( \mathcal{R}\text{ }%
-2\Lambda -F_{\mu \nu }F^{\mu \nu }\right) +\frac 1{8\pi }\int_{\partial
\mathcal{M}}d^3x\sqrt{-\gamma }\Theta (\gamma ),  \label{Actg}
\end{equation}
where $\Lambda =-3/l^2$ is the negative cosmological constant,
$F_{\mu \nu }=\partial _\mu A_\nu -\partial _\nu A_\mu $ is the
electromagnetic tensor field, and $A_\mu $ is the vector
potential. The second term $I_H$, is the action of an Abelian
Higgs system minimally coupled to gravity which can be written as

\begin{equation}
I_H=\int_{\mathcal{M}}d^4x\sqrt{-g}\mathcal{L(}\Phi ,B_\mu )=\int_{\mathcal{M%
}}d^4x\sqrt{-g}\{-\frac 12(\mathcal{D}_\mu \Phi )^{\dagger }\mathcal{D}^\mu
\Phi -\frac 1{16\pi }\mathcal{F}_{\mu \nu }\mathcal{F}^{\mu \nu }-\xi (\Phi
^{\dagger }\Phi -\eta ^2)^2\},  \label{Acth}
\end{equation}
where $\Phi$ is a complex scalar field, $\mathcal{F}_{\mu \nu }$
is the field strength associated with the field $B_\mu $, and
$\mathcal{D}_\mu =\nabla _\mu +ieB_\mu $ in which $\nabla _\mu$ is
the covariant derivative. We employ Planck units $G=\hbar =c=1$
which implies that the Planck mass is equal to unity. Defining the
real fields $X(x^\mu )$, $\omega (x^\mu )$, and$P_\mu (x^\nu )$ by
\begin{eqnarray}
\Phi (x^\mu ) &=&\eta X(x^\mu )e^{i\omega (x^\mu )},  \nonumber \\
B_\mu (x^\nu ) &=&\frac 1e[P_\mu (x^\nu )-\nabla _\mu \omega
(x^\mu )], \label{XPfield}
\end{eqnarray}
and employing a suitable choice of gauge, one can rewrite the Lagrangian (%
\ref{Acth}) and the equations of motion in terms of these fields
as
\begin{equation}
\mathcal{L(}X,P_\mu )=-\frac{\eta ^2}2(\nabla _\mu X\,\nabla ^\mu X+X^2P_\mu
P^\mu )-\frac 1{16\pi e^2}\mathcal{F}_{\mu \nu }\mathcal{F}^{\mu \nu }-\xi
\eta ^4(X^2-1)^2  \label{Lag2}
\end{equation}
\begin{eqnarray}
&& \nabla _\mu \nabla ^\mu X-XP_\mu P^\mu -4\xi \eta ^2X(X^2-1)
=0,  \nonumber
\\
&& \nabla _\mu F^{\mu \nu }-4\pi e^2\eta ^2P^\nu X^2 =0,
\label{Eqm}
\end{eqnarray}
where $\mathcal{F}^{\mu \nu }=\nabla ^\mu P^\nu -\nabla ^\nu P^\mu
$ is the field strength of the corresponding gauge field $P^\mu $.
Note that the real field $\omega $ is not itself a physical
quantity. Superficially it appears not to contain any physical
information. However, if $\omega $ is not single valued this is no
longer the case, and the resultant solutions are referred to as
vortex solutions \cite{NO}. In this case the requirement that the
$\Phi $ field be single valued implies that the line integral of
$\omega $ over any closed loop is $\pm 2\pi n$ where $n$ is an
integer. In this case the flux of electromagnetic field $\Phi
_{H\text{ \ }}$passing through such a closed loop is quantized
with quanta $2\pi /e.$

On the other hand, the static cylindrically symmetric solution of
the gravitational action (\ref{Actg}) determines the metric of a
charged black string given by \cite{Lem}
\begin{equation}
ds^2=-\Gamma dt^2+\frac {dr^2}{\Gamma} +r^2d\varphi ^2+\frac{r^2}{l^2}dz^2,
\label{met1a}
\end{equation}
\begin{equation}
A_\mu =-\frac{l\lambda }r\delta _\mu ^0,  \label{met1b}
\end{equation}
where
\begin{equation}
\Gamma =\frac{r^2}{l^2}-\frac{bl}r+\frac{\lambda ^2l^2}{r^2}.  \label{met1c}
\end{equation}
$b$ and $\lambda $ are the constant parameters of the metric which
are related to the mass and charge per unit length of the black
string by \cite {Deh3}
\begin{equation}
M=\frac b4,\hspace{1.0in}Q=\frac \lambda 2.  \label{Mas}
\end{equation}
It is worthwhile to mention that for the case of $-\infty
<z<\infty $ Eqs. (\ref{met1a})-(\ref{met1c}) describe a static
black string with cylindrical horizon. It has two inner and outer
horizons located at $r_{-}$ and $r_{+}$, provided the parameter
$b$ is greater than $b_{crit}$ given by
\begin{equation}
b_{crit}=4\times 3^{-3/4}\lambda ^{3/2}.  \label{bcrit}
\end{equation}
In the case that $b=b_{crit}$, we will have an extreme black string. \qquad

We seek a cylindrically symmetric solution for the Abelian Higgs
action (\ref{Acth}) in the background of a charged black string.
This solution can be interpreted as a skin covering the black
string (\ref{met1a})-(\ref{met1c}). Considering the static case of
winding number $N$ with the gauge choice
\begin{equation}
P_\mu (r)=(0;0,NP(r),0)  \label{Pgauge}
\end{equation}
and $X=X(r)$, and rescaling
\begin{equation}
\varkappa \rightarrow \frac \varkappa {\sqrt{\xi }\eta }  \label{rescale}
\end{equation}
where $\varkappa =r,l$, the equations of motion (\ref{Eqm}) are
\begin{eqnarray}
&&r^2l^2\Gamma X^{\prime \prime }+(4r^3-bl^3)X^{\prime
}-4l^2r^2X(X^2-1)-l^2N^2P^2(r)X(r)=0,  \label{Eqh1} \\
&&r^3l^2\Gamma P^{\prime \prime }+(2r^4+bl^3r-2\lambda
^2l^4)P^{\prime }+\alpha l^2r^3X^2(r)P(r)=0.  \label{Eqh2}
\end{eqnarray}
In the above equations (\ref{Eqh1}) and (\ref{Eqh2}), $\alpha
=4\pi e^2/\xi $ and the prime denotes a derivative with respect to
$r$. It must be noted that even in the pure flat or anti-de Sitter
spacetimes no exact analytic solutions are known for Eq.
(\ref{Eqm}). For asymptotically AdS spacetimes, one of us with
Ghezelbash and Mann showed in \cite{Deh1} that the Abelian Higgs
equations of motion in the background of anti-de Sitter spacetime
have vortex solutions with a core radius of the order of unity. In
the next section, we seek, with a numerical method, the existence
of vortex solutions for the above coupled nonlinear differential
equations.
\begin{figure}[b]
\epsfxsize=10cm \centerline{\epsffile{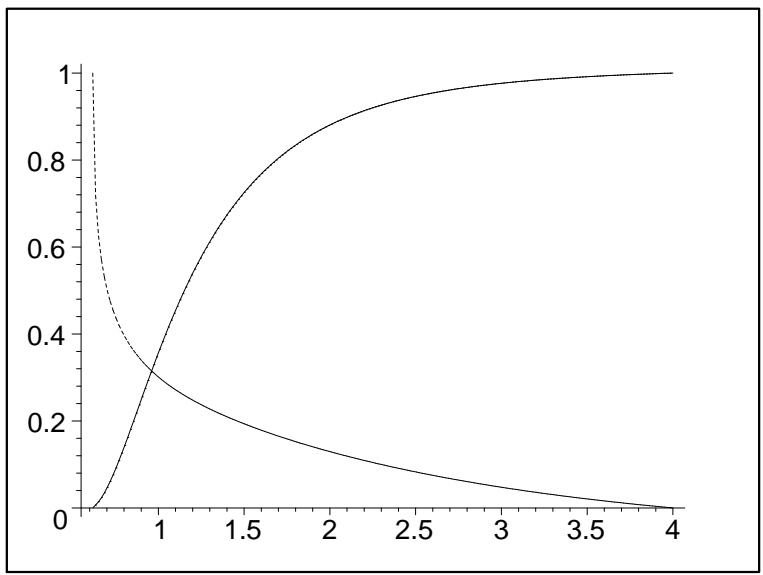}} \caption{$X(r)$
(solid) and $P(r)$ (dotted) for $l=1$, $b=0.3$, $q=0.2$, and
$N=10$.} \label{Figure1}
\end{figure}
\begin{figure}[t]
\epsfxsize=10cm \centerline{\epsffile{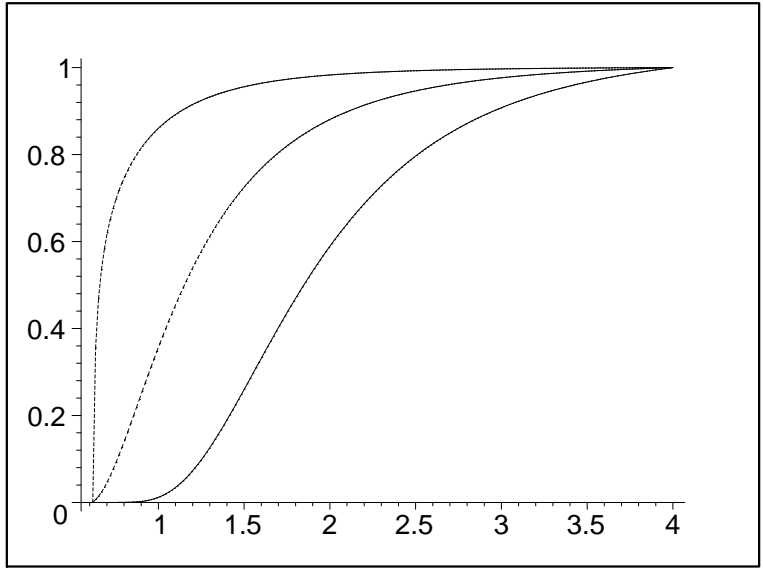}} \caption{$X(r)$ for
$l=1$, $b=0.3$, $q=0.2$, $N=1$ (dashed), $10$ (dotted), and $50$
(solid).} \label{Figure2}
\end{figure}
\section{Numerical Solutions}\label{Num}
We pay attention now to the numerical solutions of the Eqs. (\ref{Eqh1}%
) and (\ref{Eqh2}) outside the black hole horizon. First, we must
take appropriate boundary conditions. At a large distance from the
horizon, we demand that our solutions go to the \ solutions of the
vortex equations in AdS spacetime given in \cite{Deh1}. This means
that we demand $X\rightarrow 1 $ and $P\rightarrow 0$ as $r$ goes
to infinity. On the horizon, we initially take $X=0$ and $P=1.$ We
employ a grid of points $r_i$, where $r_i$ goes from $r_H$ to some
large value of $r$ ($r_\infty $) which is much greater than $r_H$.
We use the finite difference method and rewrite the nonlinear
differential equation (\ref{Eqh1}) as
\begin{equation}
a_iX_{i+1}+b_iX_{i-1}+c_iX_i=f_i,  \label{Fin}
\end{equation}
where $X_i=X(r_i).$ The coefficients $a_i,...,f_{i\text{ }}$ are
given by
\begin{eqnarray}
a_i &=&\frac{l^2r_i^2}{\left( \Delta r\right) ^2}\Gamma _i+\,\frac 1{2\Delta
r}(4\,r_i^3-bl^3),  \nonumber \\
b_i &=&\frac{l^2r_i^2}{\left( \Delta r\right) ^2}\Gamma _i-\,\frac 1{2\Delta
r}(4\,r_i^3-bl^3),  \nonumber \\
c_i &=&-2\frac{l^2r_i^2}{\left( \Delta r\right) ^2}\Gamma _i-l^2N^2P^2,
\nonumber \\
f_i &=&4\,l^2r_i^2X\left( X^2-1\right) ,  \label{Xco}
\end{eqnarray}
where $\Gamma _i$ is the value of $\Gamma $ at $r=r_i$. Equation
(\ref{Eqh2}) can be rewritten in the same form as the finite
difference equation (\ref{Fin}) by replacing $X_i$ with $P_i$ and
using the following form for the coefficients:
\begin{eqnarray}
a_i^{\prime } &=&\frac{l^2r_i^3}{\left( \Delta r\right) ^2}\Gamma _i+\,\frac
1{2\Delta r}(2r_i^4+bl^3r_i-2\lambda ^2l^4),  \nonumber \\
b_i^{\prime } &=&\frac{l^2r_i^3}{\left( \Delta r\right) ^2}\Gamma _i-\,\frac
1{2\Delta r}(2r_i^4+bl^3r_i-2\lambda ^2l^4),  \nonumber \\
c_i^{\prime } &=&r_i^3l^2\left( \frac{\Gamma _i}{\left( \Delta r\right) ^2}%
+\alpha X_i^2\right) ,  \nonumber \\
f_i^{\prime } &=&0.  \label{Pco}
\end{eqnarray}

\begin{figure}[b]
\epsfxsize=10cm \centerline{\epsffile{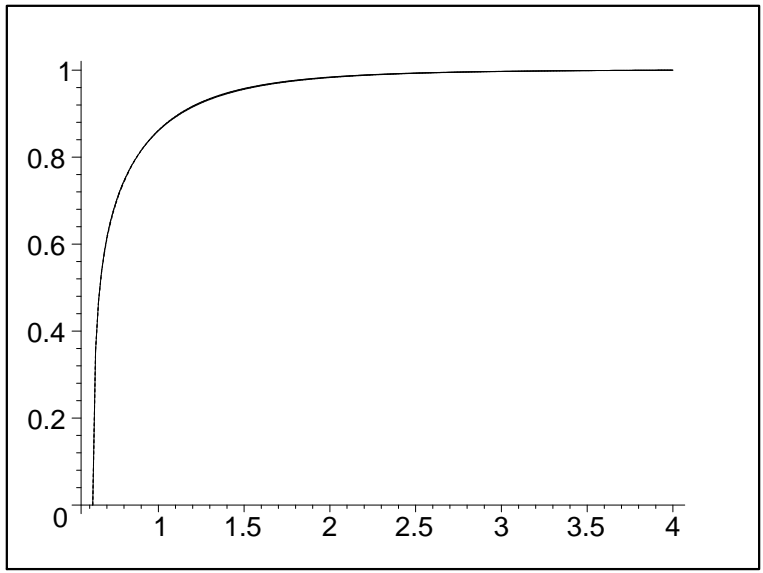}} \caption{$X(r)$ for
$N=1$, $l=1$, $r_H=0.6173$, $q=0.2$, $0.4$, and $0.5$ (three
curves overlap each other).} \label{Figure3}
\end{figure}

\begin{figure}[t]
\epsfxsize=10cm \centerline{\epsffile{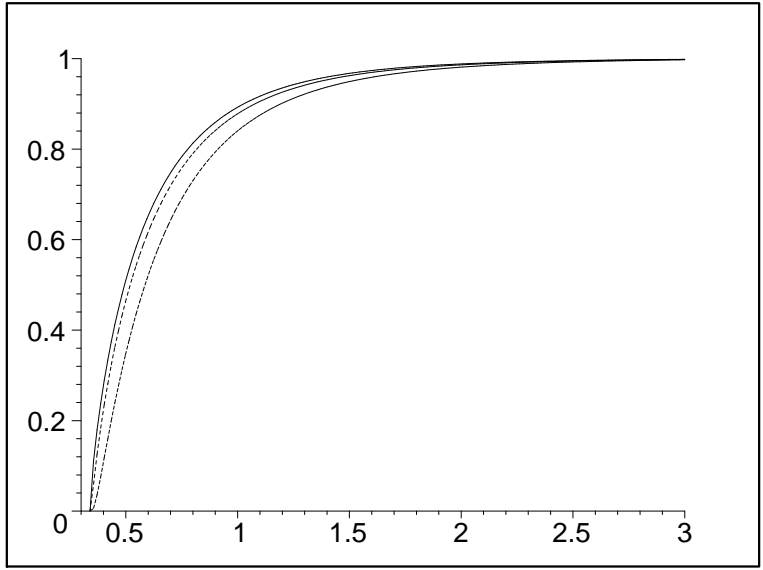}} \caption{$X(r)$
for the extreme black string $N=1$ (solid), $10$ (dotted), and
$50$ (dashed).} \label{Figure4}
\end{figure}

Now, by using the relaxation method \cite{num} for the above
finite difference equations (\ref{Xco}) and (\ref{Pco}) on a grid
which spans the domain, we calculate the numerical solutions of
$X(r)$ and $P(r)$ for different values of the metric parameters.
In all of our calculations we choose the cosmological parameter
$l=1.0 $. Some typical results of these calculations are displayed
in Figs. \ref{Figure1}-\ref{Figure4}.

Figures \ref{Figure1} shows the behavior of $X(r)$ and $P(r)$ for
$\lambda =0.2$ and $b=0.3$. The behavior of $X(r)$ for different
winding numbers $N=1$, $10$, and $50$ is displayed in Fig.
\ref{Figure2}. As one can see from these figures, as for the
asymptotically flat, dS, and AdS spacetimes considered in Refs.
\cite{Achu} and \cite {Deh1}-\cite{Ghez2}, increasing the winding
number yields a greater vortex thickness. The black string horizon
is located at $r_H=0.6173$ for Figs. \ref{Figure1} and
\ref{Figure2}. In Fig. \ref{Figure3}, $X(r)$ is plotted for
$\lambda =0.2$, $0.4$, and $0.5$, while keeping the horizon radius
constant at $r_H=0.6173$. As we see $X(r)$ is the same for these
three cases and therefore the vortex thickness does not depend on
the charge per unit length. We have also calculated the vortex
solutions for the extremal black string with cosmological
parameter $l=1.0 $, and mass and charge per unit length $b=0.157$
and
$\lambda =0.2$. In this case the horizon radius is located at $%
r_H=0.3398$. We find that the behavior of the vortex fields is the
same as in the nonextremal cases, but the thickness of the vortex
is smaller than the nonextremal case (see Figs. \ref{Figure2} and
\ref{Figure4}).
\section{Vortex Self-Gravity on the Charged Black
String}\label{Self}

We now consider the effect of the Higgs field on the charged black
string. This entails finding the solutions of the coupled
Einstein-Abelian Higgs differential equations in the charged black
string background. This is a formidable problem even for flat or
AdS$_4$ spacetimes, and no exact solutions have been found for
these spacetimes yet. However, some physical results can be
obtained by making some approximations. First, we assume that the
thickness of the skin covering the black string is much smaller
that all the other relevant length scales. Second, we assume that
the gravitational effects of the Higgs field are weak enough so
that the linearized Einstein-Abelian Higgs differential equations
are applicable.

For convenience, in this section we use the following form of the
metric for the charged black string in the presence of the Higgs
field,
\begin{equation}
ds^2=-\widetilde{A}(r)^2dt^2+\widetilde{B}(r)^2dr^2+\widetilde{C}(r)d\varphi
^2+\widetilde{D}(r)dz^2.  \label{ABCmetric}
\end{equation}
In the absence of the Higgs field, we should have $A_0(r)=\Gamma $, $%
B_0(r)=\Gamma ^{-1}$, $C_0(r)=r^2$, $D_0(r)=r^2/l^2$, yielding the
metric of the static charged black string in four dimensions.

Employing the two assumptions concerning the thickness of the
vortex and its weak gravitational field, we solve numerically the
Einstein field equations
\begin{equation}
G_{\mu \nu }-\frac 3{l^2}g_{\mu \nu }=8\pi G\mathcal{T}_{\mu \nu }
\label{Einstein}
\end{equation}
to the first order in $\varepsilon =8\pi G,$ where
$\mathcal{T}_{\mu \nu }$ is the energy-momentum tensor of the
Abelian Higgs field in the charged black string background. To the
first order of approximation, by taking $g_{\mu \nu }\simeq g_{\mu
\nu }^{(0)}+g_{\mu \nu }^{(1)}$, where $g_{\mu \nu }^{(0)} $ is
the usual charged black string metric and $\ g_{\mu \nu }^{(1)}$
is the first order correction to the metric, and writing
\begin{eqnarray}
\widetilde{A}(r) &=&A_0(r)[1+\varepsilon A(r)],  \nonumber \\
\widetilde{B}(r) &=&B_0(r)[1+\varepsilon B(r)],  \nonumber \\
\widetilde{C}(r) &=&C_0(r)[1+\varepsilon C(r)],  \nonumber \\
\widetilde{D}(r) &=&D_0(r)[1+\varepsilon D(r)],  \label{ABCexpa}
\end{eqnarray}
we obtain the corrections to the four functions $A_0(r,\theta )$,
$B_0(r,\theta ) $, $C_0(r)$ and $D_0(r,\theta )$ in Eq.
(\ref{ABCexpa}). Hence in the first approximation Eq.
(\ref{Einstein}) become

\begin{equation}
G_{\mu \nu }^{(1)}-\frac 3{l^2}g_{\mu \nu }^{(1)}=\mathcal{T}_{\mu
\nu }^{(0)}  \label{Eineq}
\end{equation}
where $\mathcal{T}_{\mu \nu }^{(0)}$ is the energy-momentum tensor
of the Higgs field in the charged black string background metric,
and $G_{\mu \nu }^{(1)}$ is the correction to the Einstein tensor
due to $g_{\mu \nu }^{(1)}$.

The rescaled components of the energy momentum tensor of the Higgs
field in the background of a charged black string are given by
\begin{eqnarray}
\mathcal{T}^t{}_t^{(0)} &=&\mathcal{T}_z^{z(0)}\,=-\frac 12\Gamma
X^{\prime
^2}-\frac{N^2}{2r^2}\Gamma P^{\prime ^2}-\frac{N^2}{2r^2}%
P^2X^2-(X^2-1)^2,  \nonumber \\
\mathcal{T}^{r(0)}{}_r &=&\frac 12\Gamma \,X^{\prime ^2}+\frac{N^2}{4\pi r^2}%
(1-2\pi )\Gamma P^{\prime ^2}-\frac{N^2}{2r^2}P^2X^2-(X^2-1)^2,  \nonumber \\
\mathcal{T}^{\phi (0)}{}_\phi  &=&\frac 12\Gamma \,X^{\prime ^2}+\frac{N^2}{%
4\pi r^2}(1-2\pi )\Gamma P^{\prime
^2}+\frac{N^2}{2r^2}P^2X^2-(X^2-1)^2, \label{stres}
\end{eqnarray}
where $X$ and $P$ are the solutions of the Higgs field.
\begin{figure}[t]
\epsfxsize=10cm \centerline{\epsffile{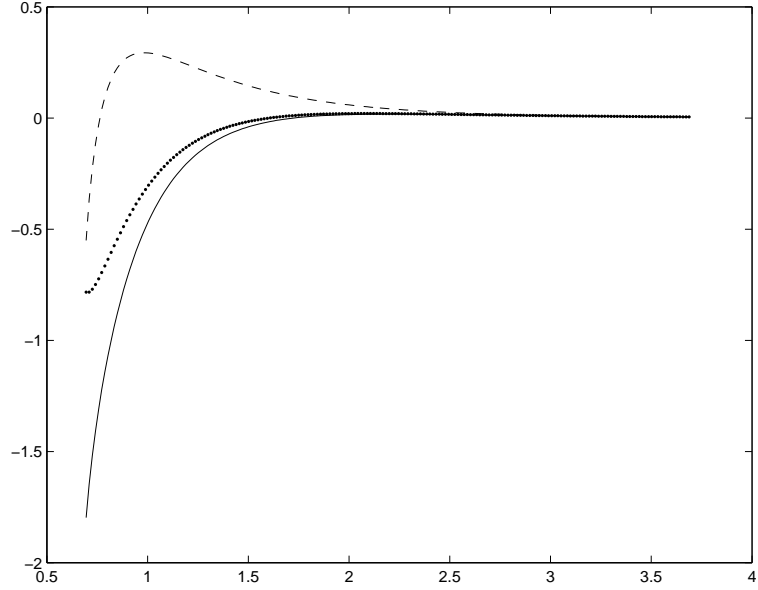}}
\caption{$\mathcal{T}_t^{t(0)}$ (solid) , $\mathcal{T}_r^{r(0)}$ (dotted), and $\mathcal{%
T}_\varphi ^{\varphi(0)}$ (dashed) for $l=1$, $b=0.3$, $q=0.2$,
and $N=1$.} \label{Figure5}
\end{figure}
\begin{figure}[b]
\epsfxsize=10cm \centerline{\epsffile{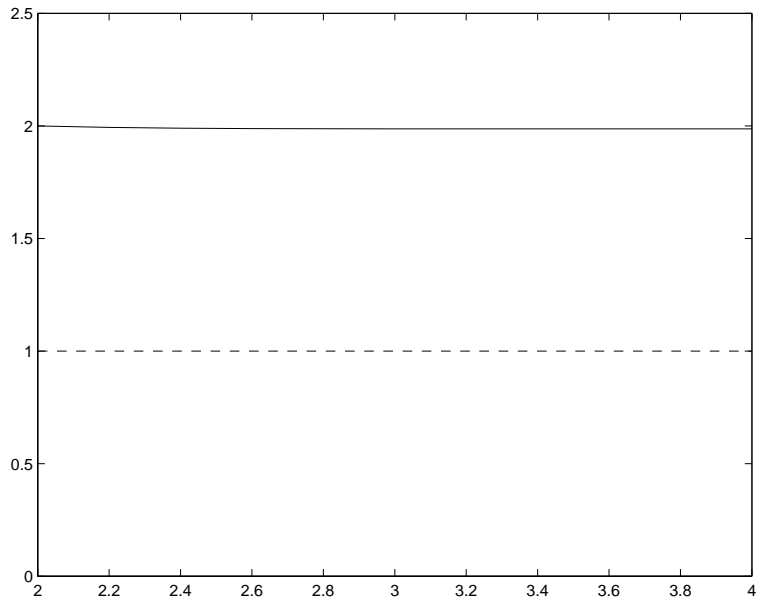}}
\caption{$A(r)$ and $B(r)$ touch the horizontal axis, $C(r)$ (dashed), and $%
D(r)$ (solid).} \label{Figure6}
\end{figure}
The Einstein equations (\ref{Eineq}) are
\begin{eqnarray}
&&\frac \Gamma 2(C^{\prime \prime }+D^{\prime \prime })-\frac \Gamma
rB^{\prime }-\frac 1r\left( \Gamma +\frac{r^2}{l^2}-\frac{bl}{4r}\right)
(C^{\prime }+D^{\prime })-\frac{\lambda ^2l^2}{r^4}A-\frac 3{l^2}B=\mathcal{T%
}^t{}_t^{(0)},  \nonumber \\
&&\frac \Gamma 2(A^{\prime \prime }+D^{\prime \prime })-\frac 1r\left(
\Gamma -\frac{r^2}{l^2}+3\frac{bl}{4r}\right) A^{\prime }-\frac 1r\left(
\frac{r^2}{l^2}-\frac{bl}{4r}\right) (B^{\prime }-2D^{\prime })+\frac{%
\lambda ^2l^2}{r^4}A-\frac 3{l^2}B=\mathcal{T}^\varphi {}_\varphi ^{(0)},
\nonumber \\
&&\frac \Gamma 2(A^{\prime \prime }+C^{\prime \prime })+\frac 1r\left(
\Gamma -\frac{r^2}{l^2}+3\frac{bl}{4r}\right) A^{\prime }-\frac 1r\left(
\frac{r^2}{l^2}-\frac{bl}{4r}\right) (B^{\prime }-2C^{\prime })+\frac{%
\lambda ^2l^2}{r^4}A-\frac 3{l^2}B=\mathcal{T}^z{}_z^{(0)},  \nonumber \\
&&\frac \Gamma rA^{\prime }+\frac 1r\left( \frac{r^2}{l^2}-\frac{bl}{4r}%
\right) (C^{\prime }+D^{\prime })-\frac{\lambda ^2l^2}{r^4}A-\frac 3{l^2}B=%
\mathcal{T}^r{}_r^{(0)}.  \label{Ein}
\end{eqnarray}
In Fig. \ref{Figure5}, the behavior of the energy-momentum tensor
components is shown. As is clear from this figure, the components
of the energy-momentum tensor rapidly go to zero outside the skin,
so the situation is like what happened in the pure AdS spacetime
\cite{Deh1}. Then solving the coupled differential equations
(\ref{Ein}) gives the behavior of the functions $A(r)$, $B(r)$,
$C(r)$, and $D(r)$ which are displayed in Fig. \ref {Figure6}. As
one can see $A(r)=B(r)=0$, and $C(r)$ and $D(r)$ are two different
constants. Hence by a redefinition of the $z$ coordinate in Eq. (\ref{ABCmetric}%
) the metric can be written as
\begin{equation}
ds^2=-\Gamma dt^2+\frac 1\Gamma dr^2+\beta ^2 r^2d\varphi ^2+\frac{r^2}{l^2}%
dz^2.  \label{metdef}
\end{equation}
The above metric describes a static charged black string metric
with a deficit angle. So, using a physical Lagrangian based model,
we have established that the presence of the Higgs field induces a
deficit angle in the charged black string metric.

\section{Conclusion}
We considered the Abelian Higgs field in the background of a
static charged black string. The numerical solutions for various
values of mass and charge parameters and winding number were
obtained. We found that for a fixed horizon radius on increasing
the winding number the vortex thickness increases. We also showed
that the thickness of the vortex is independent of the charge or
mass parameter, as long as the horizon radius and winding number
remain fixed. We also considered the extremal black string for
various winding numbers and found by numerical calculation that
the thickness of the vortex is smaller than that of the
non-extremal case. These solutions can be interpreted as stable
Abelian hair for these black strings.

Also, the effect of a thin vortex on the charged black string has
been investigated. This is done by including the self-gravity of
the thin vortex in the charged string background to the first
order of the gravitational constant. As in the case of pure AdS
\cite{Deh1}, Schwarzschild-AdS \cite{Deh2}, Kerr-AdS, and
Reissner-Nordstrom-AdS \cite{Ghez2} spacetimes, we showed that the
effect of a thin vortex on the static charged black string is to
create a deficit angle in the metric.

Other related problems such as a study of the vortex in the
charged rotating black string background and the non-Abelian
vortex solution in asymptotically AdS spacetimes, remain to be
carried out. Work on these problems is in progress.


\begin{thebibliography}{99}
\bibitem{Ruf}  R. Ruffini and J. A. Wheeler. Phys. Today \textbf{24}, 30
(1971).

\bibitem{Sud}  D. Sudarsky, Class. Quantum Grav. \textbf{12}, 579 (1995).

\bibitem{Eli}  E. Winstanley, Class. Quantum Grav\textit{.} \textbf{16}, 1963
(1999).

\bibitem{Achu}  A. Achucarro, R. Gregory, and K. Kuijken, Phys. Rev. D%
\textbf{\ 52,} 5729 (1995).

\bibitem{Torii1}  T. Torii, K. Maeda, and M. Narita, Phys. Rev\textit{. D }%
\textbf{59,} 064027 (1999).

\bibitem{Torii2}  T. Torii, K. Maeda, and M. Narita, Phys. Rev\textit{. D }%
\textbf{64,} 042116 (2001).

\bibitem{Deh1}  M. H. Dehghani, A. M. Ghezelbash, and R. B. Mann, Nucl.
Phys. \textbf{B625}, 389 (2002).

\bibitem{Ghez1}  A. M. Ghezelbash and R. B. Mann, Phys. Lett B \textbf{537},
329 (2002).

\bibitem{Cha1}  A. Chamblin, J. M. A. Ashbourn-Chamblin, R. Emparan, and A.
Sornborger, Phys. Rev. D \textbf{58}, 124014 (1998).

\bibitem{Cha2}  A. Chamblin, J. M. A. Ashbourn-Chamblin, R. Emparan, and A.
Sornborger, Phys. Rev. Lett. \textbf{80}, 4378 (1998).

\bibitem{Bon1}  F. Bonjour and R. Gregory, Phys. Rev. Lett. \textbf{81}, 5034
(1998).

\bibitem{Bon2}  F. Bonjour, R. Emparan, and R. Gregory, Phys. Rev. D. \textbf{%
59}, 84022 (1999).

\bibitem{Deh2}  M. H. Dehghani, A. M. Ghezelbash, and R. B. Mann, Phys. Rev.
D \textbf{65}, 044010 (2002).

\bibitem{Ghez2}  A. M. Ghezelbash and R. B. Mann, Phys. Rev. D \textbf{65},
124022 (2002).

\bibitem{Lem}  J. P. S. Lemos and V. T. Zanchin, Phys. Rev. D \textbf{54},
3840 (1996).

\bibitem{Deh3}  M. H. Dehghani, Phys. Rev. D \textbf{66}, 044006 (2002).

\bibitem{NO}  H. B. Nielsen and P. Olesen, Nucl. Phys. \textbf{B61}, 45 (1973).

\bibitem{num}  W. H. Press, S. A. Teukolsky, W. T. Vetterling and B. P.
Flannery, \textit{Numerical Recipes in FORTRAN} (Cambridge
University Press, Cambridge, England, 1992).
\end{thebibliography}
\end{document}